# G-Monitor: Gridbus web portal for monitoring and steering application execution on global grids

Martin Placek and Rajkumar Buyya

**Gri**d Computing and **D**istributed **S**ystems (GRIDS) Lab
Department of Computer Science and Software Engineering
The University of Melbourne, Australia
Email: mplac@students.cs.mu.oz.au and raj@cs.mu.oz.au

**Abstract.** Grids are experiencing a rapid growth in their application and along with this there is a requirement for a portal which is easy to use and scalable. We have responded to this requirement by developing an easy to use, scalable, web-based portal called G-Monitor. This paper proposes a generic architecture for a web portal into a grid environment and discusses our implementation and its application.

## 1 Introduction

Grid computing is providing us with the infrastructure to harness a heterogenous environment comprising of geographically distributed computer domains, to form a massive computing environment whereby large scale problems can be processed. For this to be achieved, Grids need to support various services [3]: security, uniform access, resource management, scheduling, application composition, computational economy and accounting [5]. The Gridbus project [7] is currently developing a "Toolkit" consisting of a suite of tools to provide these services. One of the tools within the Gridbus Toolkit is G-Monitor which aims to provide the Grid consumer with the means to control and monitor the execution of their applications within the Grid environment.

The Grid consumers need an interface to the Grid environment where they will be able to monitor, control, and steer the execution of their applications (see Figure 1). This interface must provide the user with the means to access the functionality of the Grid Resource Broker (such as Nimrod-G [5]) easily and intuitively. To provide the Gridbus toolkit with this interface the G-Monitor software package has been proposed. Although Nimrod-G provides the user with an interface with similar functionality, it is unfortunately a heavyweight client with limited functionality and scalability problems. The Nimrod-G monitor requires the user to send their X display to the machine they wish to use it on, making it very bandwidth intensive and consequently unsuitable for wide area networks. Scalability limitations within the Nimrod-G monitor arise when the user is running a large scale experiment which involves many jobs and resources.

Taking into account the problems faced by the Nimrod-G client we have developed a Web-based G-Monitor portal.

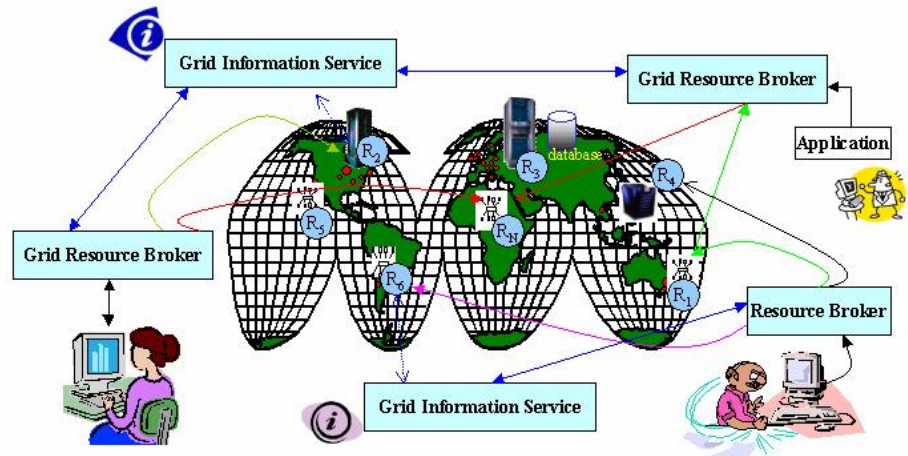

**Figure 1:** Typical Grid Computing Environment.

G-Monitor interacts with Grid Resource Broker (GRB), Nimrod-G in the current implementation, to provide the user with a GUI to the underlying Grid framework. It provides a ubiquitous interface that is easy to use, enabling the end-user to monitor and control jobs running within the Grid environment. G-Monitor is flexible enough to be run from anywhere without the need for custom client software or network overhead. G-Monitor is also scalable and therefore is able to handle thousands of nodes and jobs running in a Grid environment.

G-Monitor provides a web interface for the Nimrod-G brokering system. It allows the user to remotely monitor and control a grid system. It enables the user to:

1. **Retrieve and set QoS (quality of service) parameters, such as**
    - Deadline, Budget, Optimisation, start, stop shutdown
2. **Monitor/Control Jobs Information, such as**
    - Job name, status, remarks, grid node, Execution Time
3. **Monitor Resource status, such as**
    - Server name, Host name, Service cost, status, remarks
4. **Monitor Experiment status, such as**
    - Deadline, Budget, Job status, host status

This paper provides an insight into how G-Monitor fits into the Grid Infrastructure and how it interacts with surrounding components. We discuss G-Monitor's architecture, functionality, implementation, application and conclude by providing possible future enhancements.

## 2 Architecture

The architecture of G-Monitor and its interaction with other components in a typical Grid Environment is shown in Figure 2. G-Monitor resides on a web server, which sits between the Grid Resource Broker (GRB) and the Web Browser. The Grid Consumer uses the Web Browser to access G-Monitor's functionality. G-Monitor serves the users requests by retrieving and setting information on the GRB. The GRB manages all the Grid nodes, keeping a detailed database of information about the status of the Grid nodes whilst offering scheduling and farming out facilities.

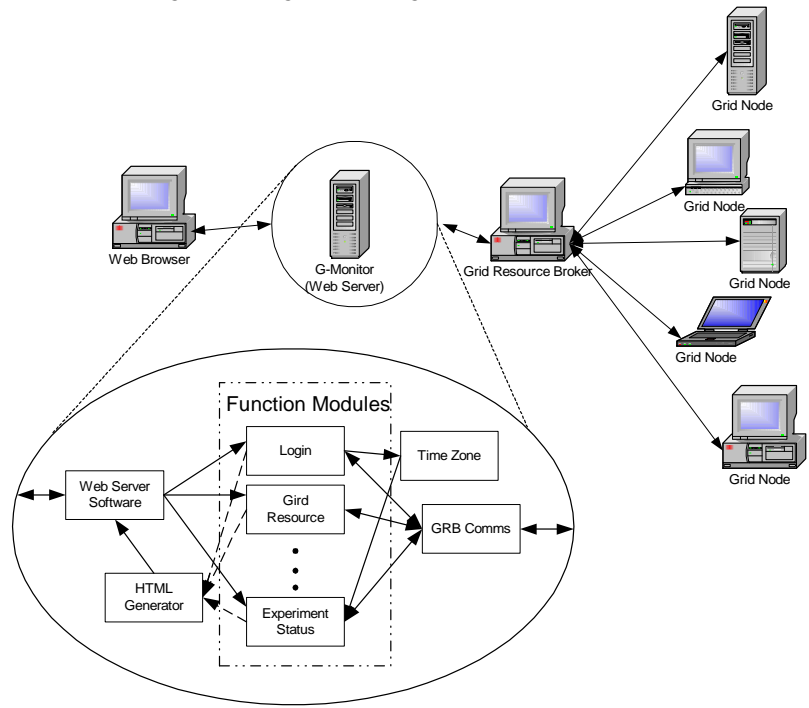

**Figure 2:** G-Monitor Architecture.

The five key components of G-monitor (as shown in magnified view in Figure 2) are:
1. **Web Server Software**: Receives http requests made by the web browser, determines which of the "Function Modules" to execute and serves the HTML generated by the "HTML Generator" back to the web browser.
2. **HTML Generator**: Receives calls from the "Function Modules" to generate HTML containing specified data. The HTML generated is then passed to the Web Server software.
3. **Function Modules**: A set of modules which provide a one-to-one mapping with the web pages offered by G-Monitor.

4. **Time Zone**: This module is called by the login script. Therefore, as the user is logging in, this module will register which timezone the user is coming from. The timezone module is then called by various "Function Modules" which require time conversion.
5. **GRB Communications**: Serves requests from the "Function Modules" by establishing a TCP socket connection with the GRB to retrieve and serve requested information.

To get a better understanding of the interaction between various components we shall walk through a work flow example. The user wishes to retrieve information on the current status of their experiment and uses the web browser to click on a link to take them to the page containing this information. The following events then occur:

1. The Web browser sends the user request off to the Web Server.
2. The Web Server software receives this request and determines which of the G-Monitor "Function Module" should be executed.
3. The "Function Module" analyses the user requests and calls upon the "GRB Comms" module.
4. The "GRB Comms" module then initiates a TCP socket to the GRB server and uses the GRB protocol (e.g., Nimrod-G job management protocol [2]) to retrieve information about the current status of the user's experiment.
5. Upon receiving the information from the GRB, the "GRB Comms" module passes this information back to the calling "Function Module".
6. The executing "Function Modules" then makes a call to the "HTML generator" which wraps the data in HTML, and serves it back to the Web Server software.
7. The Web server software then serves the HTML back to the web browser.

## 3   Implementation

The modules within G-Monitor were implemented using a combination of Perl and Javascript. The "HTML Generator", "Function Modules" and the "GRB Comms" modules were all implemented in Perl. The "Time Zone" module was implemented using a combination of Javascript and Perl. The Nimrod-G job management protocols [2] are used for interaction between the G-monitor and Nimrod resource broker.

During implementation we came across some hurdles: in particular, time zones. While testing our system locally there were no time zone differences and everything functioned well. Upon running our system from Baltimore (refer to section 5 for more detail) we found the timezone difference between the web browser and the rest of the system was causing confusion in pages which contained time related data. The solution was to create a "TimeZone" module which would be responsible for registering which timezone the user was in and making the time adjustments automatically.

## 4 Use Case Studies

G-Monitor has been deployed on a machine running an Apache Web Server in Melbourne (Australia). It has been successfully used as part of the HPC Challenge demonstration [4] at SC2002 [6] conference in Baltimore, USA (see Figure 3). This example of its use at Baltimore really highlighted how geographically distant each of the components in the Grid Infrastructure can be. G-Monitor was accessed via a web browser in Baltimore (US), while the G-Monitor and Nimrod-G modules resided in Melbourne. The grid nodes which were then used in the experiment were scattered across the world.

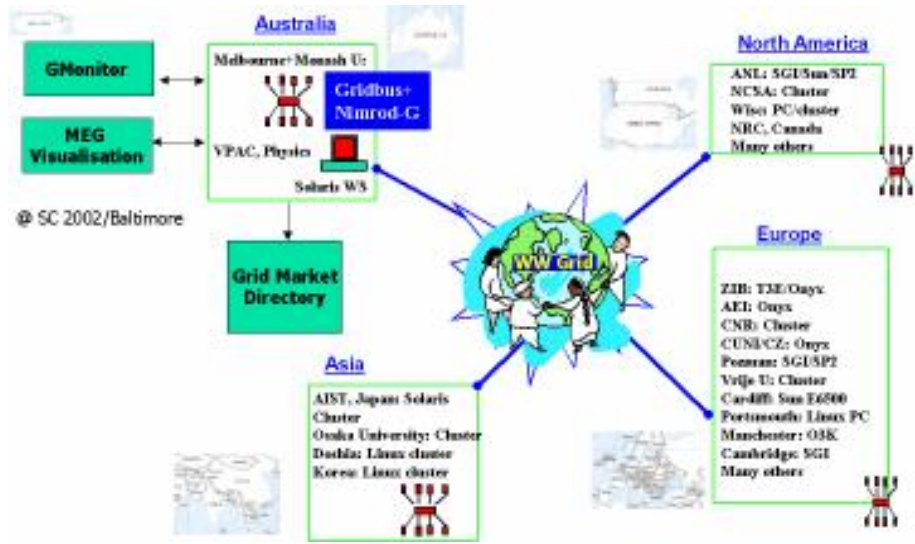

**Figure 3:** G-Monitor Usage during the HPC Challenge Demo @ SC 2002.

In the following sub-sections we shall walk through some user scenarios and get a better understanding of G-Monitor's functionality and how it can be applied in the Grid environment.

### 4.1 User sets QoS parameters

In the process of setting up an Experiment the user has the option to retrieve and set Qos Parameters using the G-Monitor (Figure 4). The user is able to set the Deadline, Budget and Optimisation preference, upon doing so the user is returned information regarding the feasibility of the settings provided. From the same page the user is also able to start, stop and shutdown their experiments.

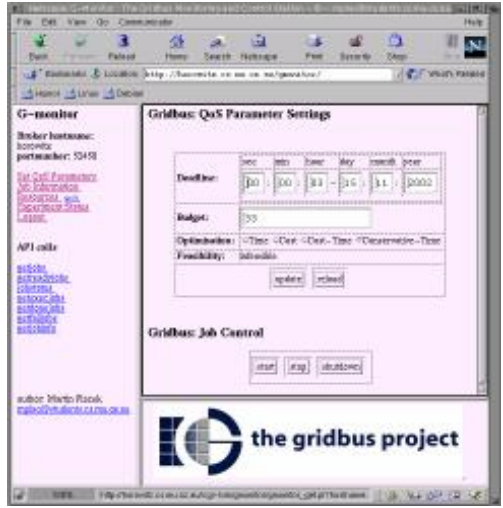

**Figure 4:** Setting QoS Parameters.

### 4.2 User experiment status

Once the user has started an experiment, they will be interested in the progress of their application execution. The users can do this by invoking the "Job Information" and "Experiment status" G-monitor Web links (see Figure 5).

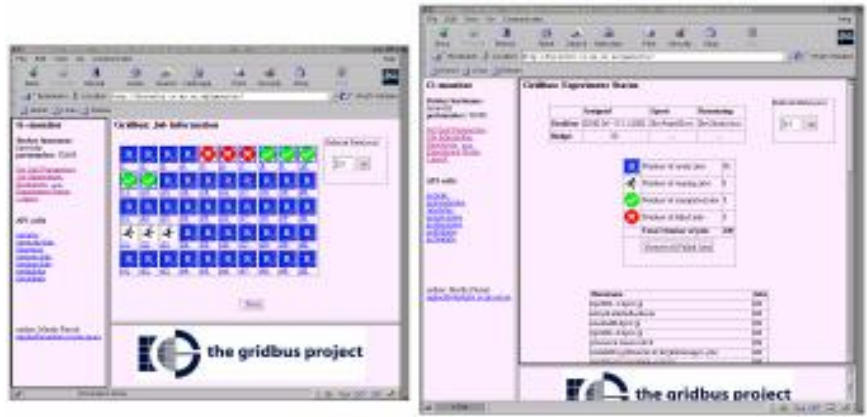

**Figure 5:** Job Information and Experiment Status.

The "Job Information" page revolves around providing information about all individual jobs. We can see that each job has a status icon associated with it, giving the user an indication whether the job is "Ready", "Running", "Completed" or "Failed". These icons are links allowing the user to drill down and retrieve more information regarding that job and allow them to have the job restarted.

The "Experiment status" page provides a summary about the status of the entire experiment. The page is broken down into three main sections. The first section provides the user with information about how their experiment is progressing with respect to: Deadline, Budget and the time remaining for completion. The second section summarises the status of all the jobs in the experiment and displays the number of current jobs that are Running, Ready, Completed or Failed. This section also provides the user with the option to restart all the failed jobs. The final section lists all the nodes in the network showing the user how many jobs have been assigned to each grid node and the number of jobs that each node has completed.

**4.3 Grid Infrastructure status**

Before running the experiment the user may want to look at what grid nodes are available and how much they cost. If the user is on a tight budget and all the cheaper resources are down, they may decide to wait for them to come back up before starting to execute their experiment. The "Resources" page (Figure 6) contains a table listing all the resources available to the user and provides information regarding the name of the server as defined within the Nimrod-G environment, the hostname, cost and its status.

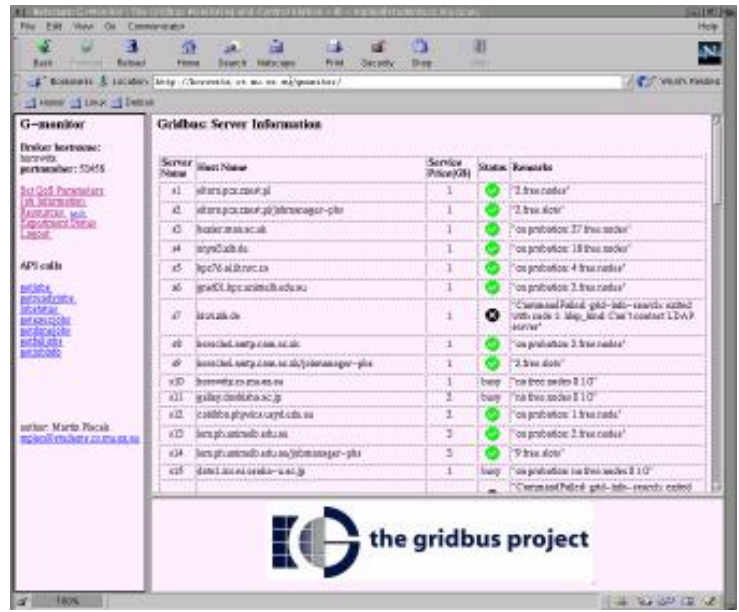

**Figure 6:** Resource Status.

## 5 Conclusion and Future work

In this paper we have identified a requirement to develop a more scalable, lightweight, easy to use GRB interface. To answer this requirement we proposed architecture for a web-based portal into a grid environment, provided an implementation (G-Monitor), and successfully demonstrated its usage in number of events including the SC 2002 HPC Challenge.

Some possible future developments for G-Monitor are as follows:

- Interface G-Monitor to other different brokerage systems eg. Entropia, UD, Parabon. Making it a universal tool which will provide a unified interface to what would otherwise be a heterogenous environment.
- Provide a graphical plot of how their experiment is progressing. eg: A plot that is updated every time delta showing its budget, number of jobs completed etc. Therefore giving the user a better idea as to the current status of the experiment.
- Provide a configuration page whereby the user is able to personalise G-Monitor for their environment. Example: Setup default Nimrod-G server, number of jobs per page, etc.

## Availability

The G-Monitor software with source code can be downloaded from the Gridbus project website:

> http://www.gridbus.org/